\documentclass[twocolumn]{jpsj2} 
\usepackage{graphicx}

\newcommand{\lsco}{La$_{2-x}$Sr$_x$CuO$_4$}
\newcommand{\ybco}{YBa$_2$Cu$_3$O$_{7-\delta}$}
\newcommand{\ybcc}{YBa$_2$Cu$_4$O$_8$}
\newcommand{\bscco}{Bi$_2$Sr$_2$CaCu$_2$O$_{8+\delta}$}
\newcommand{\pcco}{Pr$_{2-x}$Ce$_x$CuO$_{4-y}$}
\newcommand{\plcco}{Pr$_{1-x}$LaCe$_x$CuO$_{4-y}$}
\newcommand{\plccod}{Pr$_{0.89}$LaCe$_{0.11}$CuO$_{4-y}$}
\newcommand{\msr}{$\mu$SR}

\title{Strong correlation between field-induced magnetism and superconductivity
in Pr$_{0.89}$LaCe$_{0.11}$CuO$_{4}$}

\author{
R. {\sc Kadono}$^{1}$\footnote{Also at School of Mathematical and Physical Science,
The Graduate University for Advanced Studies}, 
K. {\sc Ohishi}$^{1}$, A. {\sc Koda}$^{1}$, W. {\sc Higemoto}$^{1}$,\\
K. M. {\sc Kojima}$^{2}$, M. {\sc Fujita}$^{3}$, S. {\sc Kuroshima}$^{3}$ and K. {\sc Yamada}$^{3}$}
\inst{$^{1}$Institute of Materials Structure Science, High Energy Accelerator Research Organization (KEK), Tsukuba, Ibaraki 305-0801\\
$^{2}$Department of Physics, University of Tokyo, Tokyo 113-8656, Japan\\
$^3$Institute for Chemical Research, Kyoto University, Uji, Kyoto 610-0011 }

\recdate{July 8, 2004}
\abst{
We report the local magnetic response of an electron-doped cuprate superconductor, 
\plccod\ ($T_c\simeq26$ K), studied by muon spin rotation (\msr) over a field range 
from 200 Oe to 50 kOe.  In the paramagnetic state, the muon Knight shift along 
the crystal $c$-axis ($K_\mu^z$) is proportional to the in-plane susceptibility 
$\chi_{ab}$.
More surprisingly, $K_\mu^z$ is strongly enhanced by the occurrence of 
superconductivity (below $\sim$40 kOe), 
changing its sign from positive to negative around
$\sim1$ kOe with increasing field. This finding can be understood by considering 
the weak polarization of Pr ions due to superexchange interaction with the 
antiferromagnetic Cu sublattice, which coexists with superconductivity. 
}

\kword{electron-doped cuprates, superconductivity, magnetism, $\mu$SR,\\}

\begin{document}
\sloppy
\maketitle


The role of antiferromagnetic (AF) correlation in the mechanism of 
superconductivity has been a subject of vigorous debate in the
field of high-$T_c$ cuprate superconductors since their
discovery and subsequent revelation of non-Fermi liquid
behavior in hole-doped ($p$ type) cuprates.  Despite the common
view that AF correlation is vital for the occurrence of superconductivity,
opinions are largely divided as to the ground state with which the
pair correlation is at work; namely, the one that can be mapped
onto a conventional Fermi liquid\cite{Monthoux:92}, 
or a new state of the matter consisting of 
``spinons/holons"\cite{Suzumura:88} or a texture of self-organized carriers 
such as ``stripes"\cite{Zaanen:98}.
Recent observation of field-induced magnetism in $p$ type cuprates has  
shed new light on this issue, where the recovery of the quasistatic AF state 
under a moderate magnetic field (a few Tesla) is suggested in the
flux line lattice (FLL) state of \lsco\ (LSCO)\cite{Katano:00,Lake:01,Lake:02}, 
\ybco\ (YBCO)\cite{Mitrovic:01}, \ybcc\ (Y$_{1248}$)\cite{Kakuyanagi:02}, 
and \bscco\ (BSCCO)\cite{Hoffman:02}.  Since the superconducting 
order parameter is locally suppressed in the vortex cores, there is a possibility
that these findings may be a manifestation of the electronic ground state 
nucleated in the normal cores.
However, while all of the observed effects are 
attributed to the AF (or quasistatic stripe) phase localized
in the vortex cores, the evidence for the ``AF core" remains elusive; 
the results in LSCO are all from neutron diffraction measurements which are not
sensitive to the local structure over such a large length scale
(vortices with a core size of $\sim10^2$ \AA, separated
by $10^2$--$10^3$ \AA), those in YBCO and Y$_{1248}$ are
based on NMR measurements where the evidence is an enhancement of 
spin-lattice relaxation near the core rather than a well-defined shift,
and that for BSCCO is from STM where the measured quantity is subject to the 
strong anisotropy of tunnel conductance. One attempt to directly observe
modified field distribution near the vortex cores in YBCO ($\delta=0.5$)
by the muon spin rotation
(\msr) technique yielded only a marginal sign of such modulation\cite{Miller:02}.

Meanwhile, there is increasing evidence that the ground state of
electron-doped ($n$ type) cuprates may be understood within the framework of a
conventional Fermi liquid.  The normal state of $n$ type cuprates exhibits a quadratic 
temperature dependence ($\propto T^2$) of the resistivity 
in contrast to the linear dependence seen in $p$ type cuprates.
A recent NMR study of an $n$ type cuprate, \plcco\ (PLCCO, $x=0.09$), has 
demonstrated that the Korringa law is restored upon the removal of 
superconductivity by applying
an external magnetic field above the upper critical field 
($H_{c2}\simeq50$ kOe)\cite{Zheng:03}.  These are typical features found in
the Fermi liquid ground state. Moreover, the spin dynamics
in the superconducting state of an $n$ type cuprate exhibits 
commensurate spin fluctuation\cite{Yamada:03},
whereas incommensurate spin fluctuation is commonly found in $p$ type cuprates.
In these circumstances, the field-induced AF state recently observed 
in Nd$_{1.85}$Ce$_{0.15}$CuO$_4$ is attracting much interest\cite{Kang:03,Matsuura:04}; 
it would be crucially important to examine whether or not such AF phase is localized 
within the vortex cores.

Recently, a much stranger external field  effect in 
\pcco\ (PCCO) has been observed by \msr,
where large copper moments ($\sim0.4$ $\mu_B$) are reported to have been
induced by a magnetic field as small as 90 Oe\cite{Sonier:03}.
In this paper, we report \msr\ measurements on the single crystalline $n$ type
cuprate \plccod\  ($T_c\simeq26$ K), which is relatively close to the 
AF phase ($x\le0.1$)\cite{Fujita:03}.
Compared with PCCO, we can obtain single crystals of PLCCO in much larger dimensions,
which allows systematic studies by various experimental techniques including neutron 
diffraction\cite{Fujita:03-2}.
We show that the muon Knight shift, $K_\mu^z$, under an external magnetic field
parallel to the $c$-axis is significantly influenced by the in-plane susceptibility, 
$\chi_{ab}$, strongly suggesting the presence of unconventional hyperfine interaction
involving a non-diagonal Fermi contact-type term between muons and Pr ions.
Moreover, an additional shift of the frequency (corresponding to $\sim10^1$ Oe) 
is induced spontaneously below $\sim T_c$ at lower fields ($<40$ kOe) 
with either a positive or negative sign, depending on the magnitude of the applied 
magnetic field.  
This indicates that the in-plane polarization of Pr ions is clearly enhanced in
the superconducting phase, strongly suggesting the influence of CuO$_2$ planes
on the behavior of Pr moments.  Considering the field-induced weak antiferromagnetism
of CuO$_2$ planes in PLCCO\cite{Fujita:03-2}, we discuss the superexchange interaction 
between Pr and Cu moments as an origin of Pr moment polarization.

\begin{figure}[hbt]
\includegraphics[width=0.9\linewidth]{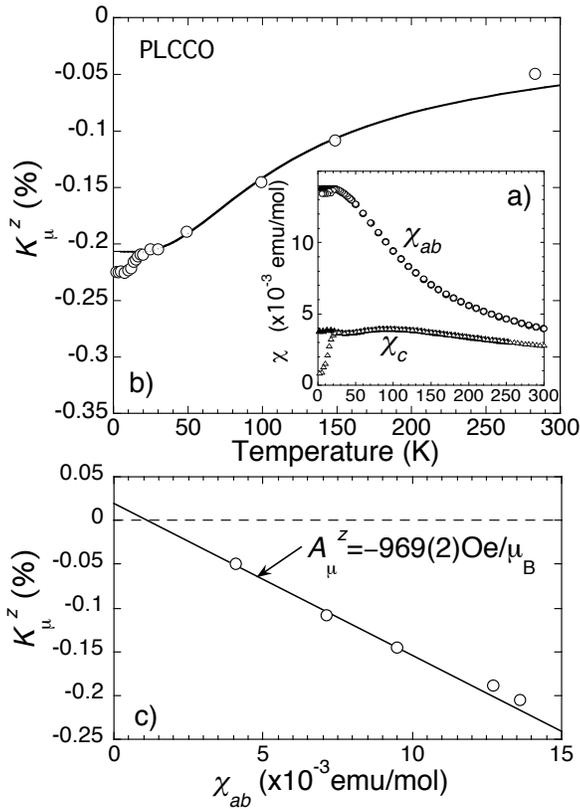}%
\caption{\label{fig1} (a) Temperature dependence of 
magnetic susceptibility at $H$=20 kOe (open symbols) and 50 kOe (filled symbols) 
applied parallel  ($\chi_{c}$) and perpendicular ($\chi_{ab}$) to the
$c$-axis.  (b) The muon Knight shift with $H=20$ kOe parallel to the $c$-axis, 
where the solid curve in (b) is proportional
to $\chi_{ab}$.  (c) The $K$-chi plot (with $\chi=\chi_{ab}$) 
for the Knight shift shown in (b). }
\end{figure}

A large single crystal of \plccod\ ($y\simeq0.02$) was prepared by the 
traveling-solvent float zone method, the details of which have already been published 
elsewhere\cite{Fujita:03}. A large volume fraction 
and the sharp onset of  Meissner diamagnetism at $T_c$ 
(see below) demonstrated the high quality of the specimen.  
The \msr\ measurements were carried out on the M15 beamline at TRIUMF, Canada.
A slab of \plccod\ crystal (measuring about 
5 mm$\times$8 mm$\times$0.5 mm) with the $c$-axis perpendicular to the 
plane was loaded onto a He gas-flow cryostat and a magnetic field (${\bf H}=(0,0,H_z)$) 
was applied  parallel to the $c$-axis (where $z\parallel c$). 
In a transverse field (TF) geometry, the initial 
muon polarization was perpendicular to the $c$-axis so that the muon probed
the local field $B_z$ by spin precession at a frequency $\gamma_\mu B_z$
(with  $\gamma_\mu=13.553$ MHz/kOe being the muon gyromagnetic ratio).
Detailed zero-field (ZF) \msr\ measurements on the same specimen 
with varied oxygen depletion indicated a weak random magnetism similar
to PCCO, which is identified as being due to the small Pr moments\cite{Kadono:03}.

The muon hyperfine parameter $A_\mu^z$ is deduced from a 
comparison between magnetic susceptibility $\chi$ and 
the muon Knight shift in the normal state; 
their relation in rare-earth metallic compounds is generally expressed as
\begin{equation}
K_\mu^z\simeq K_0+(A_{c}+\sum_iA^{zz}_i)\chi_{c},\label{kzc}
\end{equation}
where $K_0$ and $A_c$ denote the respective contributions from the 
$T$-independent Pauli paramagnetism and
from the polarization of conduction electrons by the Rudermann-Kittel-Kasuya-Yoshida 
(RKKY) interaction, $\chi_c$ being the susceptibility for the normal directions, 
and $A^{zz}_i$ is the relevant component of the dipole tensor 
$A^{\alpha\beta}_i=(3r^\alpha_i r^\beta_i/r^2_i-
\delta_{\alpha\beta})/r^3_i$ calculated for the nearby $i$-th Pr ions
at a distance ${\bf r}_i$ from the muon. 
As shown in Fig.~\ref{fig1}a, PLCCO exhibits a large anisotropy of $\chi$ 
between the in-plane ($\chi_{ab}$) and normal ($\chi_c$) directions, 
where $\chi_{ab}$ exhibits a significant increase with decreasing temperature 
while $\chi_c$ remains almost unchanged.   The corresponding muon Knight
shift versus temperature is shown in Fig.~\ref{fig1}b; it is clear 
in Figs.~\ref{fig1}a and \ref{fig1}b that $K_\mu^z$ is mostly 
proportional to $\chi_{ab}$, which is in a stark contrast to  Eq.~(\ref{kzc}).
Moreover, it is inferred from Fourier analysis of the
TF-\msr\ spectra that there are two satellite peaks symmetrical about the central 
peak with about a half intensity
whose splitting becomes large enough to be resolved 
above $\sim$25 kOe\cite{Kadono:03}.  
Provided that the rare-earth sites are randomly occupied by Pr and 
La ions, the observed spectral pattern suggests a binomial
distribution of hyperfine coupling constants.  

\begin{figure}[hbt]
\includegraphics[width=0.9\linewidth]{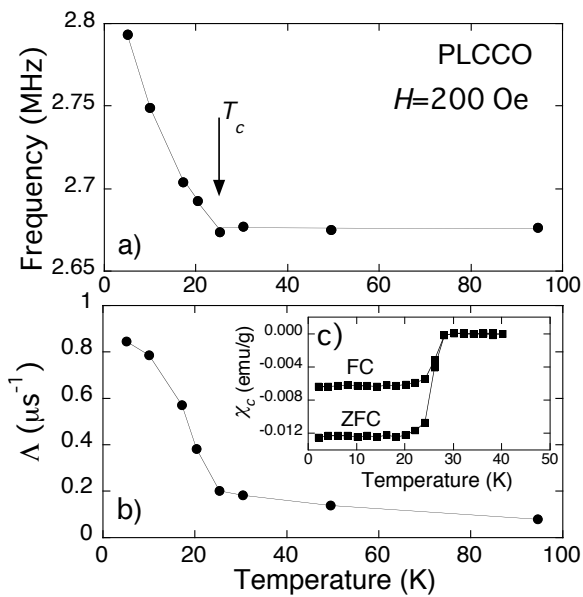}%
\caption{\label{fig2} Temperature dependence of the muon precession frequency 
(a) and transverse spin relaxation rate (b) at $H=200$ Oe. The inset (c)
shows the Meissner effect observed in bulk susceptibility at 10 Oe under FC and 
ZFC conditions. The solid lines are a visual guide only.}
\end{figure}

Considering the earlier reports that the muon site is crystallographically unique 
and located near the oxygen atoms 
midway between the CuO$_2$ planes  
($z=0.25$ or 0.75 in the unit cell)\cite{Luke:90,Le:90}, we carried out extensive 
simulations of $K_\mu^z$ using Eq.~(\ref{kzc}). 
As a result, we found that there is no possibility for Eq.~(\ref{kzc}) 
to reproduce the well-defined three-frequency structure 
with binomial intensity distribution observed at higher 
fields.  This is also obvious from the fact that the temperature dependence of 
$K_\mu^z$ is quite different from that of $\chi_c$.
Thus,  we need to introduce a hypothetical non-diagonal
hyperfine interaction, $A_{f}$, to account for the observed result; namely,
\begin{equation}
K_\mu^z\simeq K_0+A_{f}\chi_{ab}.\label{kzab}
\end{equation}
Our result in Fig.~\ref{fig1}b indicates that the term proportional to 
$\chi_c$ is negligible.  (Note that the dipolar tensor $\sum A^{zz}_i$ actually takes
null value when muons are sitting exactly at the center between two rare-earth ions.)
The $K$-chi plot for $\chi_{ab}$ in Fig.~\ref{fig1}c exhibits
a linear relation with a small offset near the origin, from which we obtain
$A_{f}=A_\mu^z=-969(2)$ Oe/$\mu_B$ for the central peak, and
$-401(2)$ Oe/$\mu_B$  and $-1551(2)$ Oe/$\mu_B$ for the satellites
assuming a common $K_0$ (=201(3) ppm).  
 Although the origin of $A_{f}$ is not clear at this stage, it is most probable
that a Fermi contact-type (scalar) interaction (a ``variation" of $A_c$)
is responsible for the mechanism explaining the binomial distribution of the spectral
weight. 

\begin{figure}[hbt]
\includegraphics[width=0.85\linewidth]{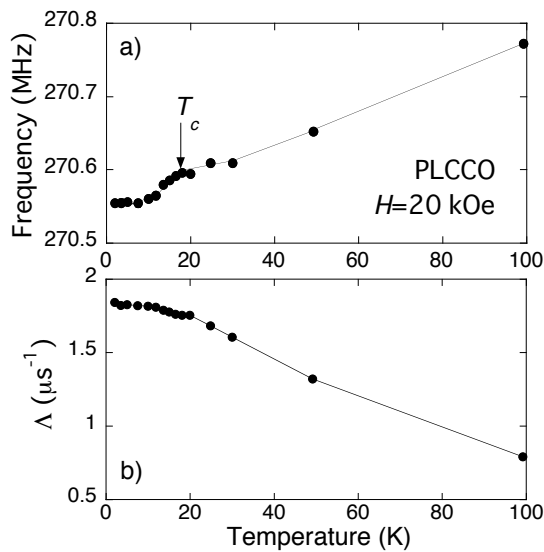}%
\caption{\label{fig3}
Temperature dependence of the muon precession frequency 
(a) and transverse spin relaxation rate (b) at $H=20$ kOe
(where $T_c\simeq 18$ K). The solid lines are a visual guide only.}
\end{figure}

At the superconducting transition, the specimen exhibits a further peculiar
magnetic response.  As shown in Fig.~\ref{fig2},
a large {\sl positive} shift of the frequency and associated increase of
the muon spin relaxation rate ($\Lambda$)  is observed at $H=200$ Oe. 
While the increase of $\Lambda$ is understood by
considering the inhomogeneous magnetic field distribution due to
the FLL formation, the direction of the frequency shift is apparently
opposite to that expected for diamagnetism in the FLL state.
As shown in Fig.~\ref{fig2}c, the possibility of attributing the 
observed positive shift to the so-called paramagnetic 
Meissner effect is ruled out by the magnetic response observed in both the ZFC 
and FC magnetization. This result is similar to what has been observed in 
PCCO\cite{Sonier:03}.  In Fig.~\ref{fig3}, on the other hand, 
an enhancement of the frequency shift in the {\sl negative} direction 
is observed below $T_c$ at 20 kOe.

\begin{figure}[hbt]
\includegraphics[width=0.9\linewidth]{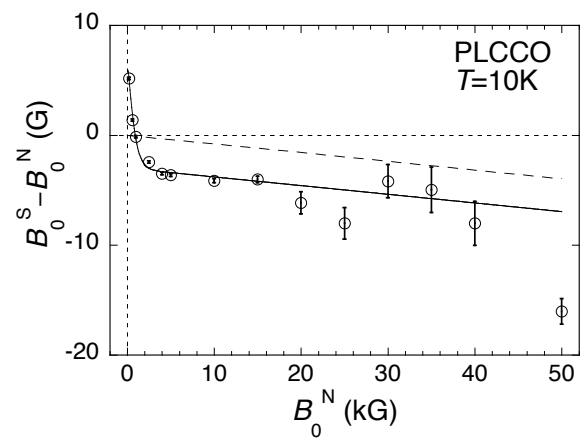}%
\caption{\label{fig4}Magnetic field dependence of the additional 
shift in the superconducting state,
where $B^S_0$ and $B^N_0$ correspond to the internal
field at 10 K and 30--40 K, respectively. The solid curve is a ``guide for the eye"
fitting result by a phenomenological model assuming the flip of an internal field 
(with the dashed line representing the contribution of residual demagnetization).}
\end{figure}

The field dependence of this additional shift, 
$\Delta B_z=B^S_0-B^N_0$ (frequencies divided by $\gamma_\mu$), is shown in 
Fig.~\ref{fig4}, where $\Delta B_z$ exhibits a steep decrease with increasing
field to {\sl change its sign} to negative above $\sim1$ kOe.  
The strong field dependence at such low fields indicates that the shift of the Van Hove
singularity due to the FLL formation is negligible over the entire field range; this is also
consistent with a large magnetic penetration depth ($\ge3400$ \AA) reported for PCCO.
While the observed tendency is close to that in PCCO ($H\le 2$ kOe)\cite{Sonier:03}, 
the change in the sign of $\Delta B_z$ suggests the flip of $A^z_\mu$ 
due to the external field. 
(Unfortunately, $K^z_\mu$ in the normal state is too small to measure at such low fields.)
Meanwhile, $\Delta B_z$ is only weakly
dependent on the field for 2--40 KOe, above which it exhibits
a trend to increase further in the negative direction.  Such a behavior 
cannot be explained by the simple bulk demagnetization effect.
More interestingly, these features above $\sim$2 kOe (including a further increase of
$\Delta B_z$ above $\sim40$ kOe) are quite similar to those of field-induced 
moments detected by neutron diffraction\cite{Fujita:03-2}.
The possibility to attribute the field-induced effect to impurity phases
(Pr,Ce,La)$_2$O$_3$ has been ruled out by the recent neutron experiment, as
they found no such effect in those phases below 70 kOe\cite{Matsuura:04}.

Since the emergence of $\Delta B_z$ is strongly correlated with the occurrence
of superconductivity below $\sim40$ kOe, the origin of $\Delta B_z$ 
must be in close relation 
with the electronic state of CuO$_2$ planes. However, it has been
demonstrated by very recent
neutron diffraction measurements on the present specimen that
the moment size of Cu ions induced by the external field is far smaller
than that found in the AF phase of \plcco\ ($x<0.1$)
with a quite non-linear dependence on the field; it remains almost constant over
a range from 0 to $\sim40$ kOe\cite{Fujita:03-2}.
Considering the hyperfine coupling between Cu ions and
muons at the relevant site (which is usually dominated by magnetic
dipolar interaction), it is unlikely that such small Cu moments directly 
contribute to $\Delta B_z$, in contrast to what has been suggested for the case of 
PCCO\cite{Sonier:03}. 
On the other hand, there is a strong superexchange coupling between
Cu and Pr ions as inferred from neutron diffraction studies
in $R_2$CuO$_4$\cite{Matsuda:90}.  For example, in 
Pr$_2$CuO$_4$ about 0.08 $\mu_B$ of Pr moments are induced
by Cu moments with 0.4 $\mu_B$\cite{Matsuda:90}, with a 
non-collinear spin structure in the $ab$ planes for both 
sublattices\cite{Skanthakumar:93}.
In a mean-field treatment, the Cu ions exert an effective magnetic field
on the Pr ions so that the moment size of the Pr ions is given by 
\begin{equation}
\langle M_{\rm Pr}\rangle\sim \chi_{ab}J\langle M_{\rm Cu}(H_z)\rangle,
\end{equation}
where $J$ is the Cu$^{2+}$-Pr interaction energy and 
$\langle M_{\rm Cu}(H_z)\rangle$ is the field-induced Cu moments\cite{Matsuda:90}.
Thus, we can expect an additional field 
$\Delta B_z\simeq A_{f}\langle M_{\rm Pr}\rangle$,
which is induced by the polarized Cu ions in the superconducting phase.
Because of the non-diagonal hyperfine coupling $A_{f}$ 
between muons and Pr ions, a small {\sl in-plane} polarization of 
Pr ions can lead to a sizable magnitude of hyperfine field along the 
$c$-axis on the nearby muons.
Our estimation indicates that about 0.01 $\mu_B$ of Pr moments, 
which may be induced by 0.05 $\mu_B$ of Cu moments, 
is enough to account for the amplitude of $\Delta B_z$.

It is interesting to note that the strong influence of the superconducting phase is observed
at as low an external field as $10^2$ Oe, where the density of magnetic
vortices is very small (their distance being $\sim 4\times10^3$ \AA).  Considering
that most of the implanted muons are probing the sample {\sl outside}
the vortex cores in this low field range,  we can conclude that the origin of
the enhanced muon Knight shift associated with the superconducting phase is not 
confined in the vortex cores.  
Thus, it is highly presumable that the small polarization of Pr ions (and of Cu ions)
is present in the entire volume of the specimen under an external field, 
as suggested by the neutron diffraction.  This might be in favor of the 
quantum critical point scenario in understanding the competition between AF 
and superconducting phases\cite{Sachdev:03}.

Finally, we note that additional work is clearly needed to elucidate the origin of 
unconventional hyperfine coupling between muon and Pr ions, 
including that on the possible role of muon itself as a source of perturbation.

In summary, we have shown that the muon Knight shift, which is anomalously
proportional to the  {\sl in-plane} susceptibility (perpendicular to the field
direction), is strongly enhanced by the occurrence of superconductivity, 
changing its sign from positive to negative around $\sim1$ kOe with increasing field. 
This result can be understood by considering weak in-plane polarization of Pr ions,
which are polarized by field-induced Cu moments via superexchange interaction.
It also suggests strongly that there is a uniform weak polarization of 
Cu moments induced by an external field as low as 10$^2$ Oe in 
the superconducting phase of \plccod, irrespective of vortex cores
in the mixed state.  This extraordinary sensitivity of CuO$_2$ planes
to a magnetic field will provide a strong criterion for identifying the true
electronic ground state of $n$ type cuprates.

We wish to thank the staff of TRIUMF for their technical support
during the experiment. This work was partially supported by a Grant-in-Aid
for Scientific Research on Priority Areas and a Grant-in-Aid 
for Creative Scientific Research by the Ministry of Education, Culture,
Sports, Science and Technology, Japan




\end{document}